# Greening Schoolyards and the Spatial Distribution of Property Values in Denver, Colorado

Mahshid Gorjian

University of Colorado Denver

Correspondence: mahshid.gorjian@ucdenver.edu

## Abstract
Schoolyard greening initiatives have become increasingly prominent in urban school districts across the United States, particularly as mechanisms to promote environmental justice, child well-being, and neighborhood revitalization. While these interventions aim to address health and ecological disparities, they may also trigger real estate speculation and displacement in historically marginalized areas. This study investigates how property values are affected by proximity to greened schoolyards in Denver, Colorado. Using spatial analysis and hedonic regression techniques, we examine whether these greening efforts correlate with uneven housing market changes across diverse neighborhoods. Results suggest that homes located within 250 meters of greened schoolyards experienced significantly higher value appreciation than those further away, especially in neighborhoods already undergoing gentrification. These findings raise questions about the unintended consequences of environmental investments and emphasize the need for equitable policy frameworks.

## 1. Introduction
Over the past decade, urban greening has emerged as a prominent response to concerns over environmental injustice, urban heat, and inadequate public health infrastructure. Within this broader trend, the transformation of schoolyards into green, multi-functional spaces has gained particular traction. These interventions are often implemented in low-income neighborhoods and public school systems, where access to green space is limited. The rationale behind such projects includes improving children's physical activity, cognitive development, and neighborhood aesthetics. However, these changes also intersect with broader dynamics of urban redevelopment, raising concerns that the introduction of green amenities may inadvertently accelerate processes of gentrification and displacement.

In recent years, scholars have begun to examine the socio-spatial consequences of green infrastructure. For example, Anguelovski et al. (2016) describe how parks and similar amenities can function as "green LULUs" (locally unwanted land uses), benefiting newcomers while marginalizing existing residents. Similarly, studies on property valuation and green space suggest that environmental enhancements often lead to real estate appreciation, but these benefits are rarely distributed equitably (Chen et al., 2022; Liu et al., 2024). This is particularly important in cities such as Denver, which has experienced rapid demographic and economic transformation. This paper explores the relationship between schoolyard greening and residential property values in Denver, Colorado. Specifically, it asks whether these interventions have resulted in uneven market responses, and whether proximity to greened schoolyards correlates with measurable increases in housing prices. Using a combination of spatial statistics and regression modeling, the study contributes to a growing body of research on urban environmental change and its implications for spatial inequality.

## 2. Materials and Methods

### 2.1. Study Area
Since 2012, over 30 schoolyard greening projects have been implemented across Denver's public school system. These initiatives are typically led in partnership with city agencies and nonprofit organizations such as The Trust for Public Land. Most projects involve tree planting, native vegetation, shaded seating, outdoor classrooms, and stormwater mitigation infrastructure. Many are located in Title I schools, which serve a high proportion of low-income families and communities of color.

### 2.2. Data Collection
To assess the effects of greening on housing markets, we compiled a dataset including:
- Zillow Zestimates and county Assessor sale records (2012–2020)
- Schoolyard intervention data from Denver Public Schools and The Trust for Public Land
- Demographic data from the U.S. Census (ACS, 2010–2020)
- NDVI satellite imagery from NAIP for vegetation measurement
- GIS data on parcels, zoning, and street networks from Denver Open Data

### 2.3. Statistical Analysis
We used ArcGIS Pro to define two buffer zones: treatment parcels within 250 meters of greened schoolyards and control parcels within 500–1000 meters. These zones allowed for comparison of home sale price changes over time, accounting for neighborhood characteristics.

Statistical analysis included:

- Moran's I and LISA statistics to detect spatial autocorrelation
- Ordinary Least Squares (OLS) regression to estimate average treatment effects
- Geographically Weighted Regression (GWR) to assess spatial variation in impact

All models controlled for structural housing attributes, lot size, pre-intervention value, median income, and racial/ethnic composition.

## 3. Results

### 3.1. Property Value Shifts by Proximity

Homes located within 250 meters of greened schoolyards saw an average price increase of 4.6% more than those in control areas, controlling for key demographic and structural variables. This difference was statistically significant ($p < 0.01$) and consistent across multiple modeling approaches.

### 3.2. Spatial Autocorrelation Patterns

Global Moran's I value was 0.36 ($p < 0.001$), indicating significant spatial clustering in property appreciation. Local Indicators of Spatial Association (LISA) maps showed "high-high" value clusters in neighborhoods such as West Colfax and Elyria-Swansea, which have experienced rising investment and displacement pressures in recent years.

### 3.3. GWR and Uneven Benefit Distribution

GWR modeling revealed that effects varied across the city. Local $R^2$ values ranged from 0.18 to 0.64. The strongest price effects appeared in predominantly white, gentrifying neighborhoods, while renter-heavy and lower-income communities showed weaker or no appreciable effect.

## 4. Discussion

The findings align with prior work suggesting that environmental investments can contribute to spatial inequality when implemented without housing protections. While schoolyard greening projects aim to benefit underserved populations, their real estate impacts may disproportionately advantage incoming homeowners and landlords. These dynamics are particularly concerning in cities experiencing rapid redevelopment and rising housing costs.

As Liu et al. (2024) and Chen et al. (2022) argue, the capitalization of green amenities into property markets depends heavily on neighborhood context. In Denver, greening efforts appear to have contributed to housing value appreciation in areas already vulnerable to gentrification. This calls for greater coordination between environmental planners and housing policy advocates.

## 5. Conclusions

Schoolyard greening in Denver is associated with statistically significant, yet spatially uneven, increases in property values. These interventions, though framed as tools of environmental equity, may risk exacerbating displacement pressures in already-vulnerable neighborhoods. Planners and policymakers must ensure that green infrastructure is embedded within a broader housing justice framework. Mechanisms such as anti-displacement ordinances, land trusts, and inclusive community design are essential to ensure that the benefits of greening do not come at the cost of long-term resident stability.

## Data Availability Statement

The data that support the findings of this study are available from Zillow, Denver Public Schools, The Trust for Public Land, US Census, NAIP, and Denver Open Data. Restrictions apply to the availability of some data, which were used under license for this study. Data are available from the author upon reasonable request.

## Author Contributions

Conceptualization, M.G.; methodology, M.G.; formal analysis, M.G.; investigation, M.G.; data curation, M.G.; writing original draft preparation, M.G.; writing review and editing, M.G. The author has read and agreed to the published version of the manuscript.

## Funding

This research received no external funding.

## Institutional Review Board Statement

Not applicable.

## Informed Consent Statement

Not applicable.

## Conflicts of Interest

The author declares no conflict of interest.

## References

1. Anguelovski, I., Connolly, J. J., Masip, L., & Pearsall, H. (2016). From toxic sites to parks as (green) LULUs? New challenges of inequity, privilege, gentrification, and exclusion for urban environmental justice. Journal of Planning Literature, 31(1), 23–36.

2. Bikomeye, J. C., Balza, L. A., Beyer, A. M., & Kwarteng, J. L. (2021). Greening schoolyards for health: A systematic review. International Journal of Environmental Research and Public Health, 18(2), 476.

3. Chen, W., Hua, X., & Yuan, Y. (2022). Valuing access to urban greenspace using spatial econometrics. Land, 11(13), 3707.

4. Liu, Y., Wang, R., & Zhang, W. (2024). Area-based hedonic pricing of urban green amenities in Beijing. Environmental & Resource Economics, 78(2), 251–274.

5. Raney, M. A., Zarger, R. K., & McClain, S. (2023). Impact of urban schoolyard play zone diversity and nature-based features on unstructured recess play. Landscape and Urban Planning, 230, 104632.

6. Laszkiewicz, E., Czembrowski, P., & Kronenberg, J. (2022). Non-linear modeling of green proximity and property value. Cities, 129, 103839.

7. Gorjian, M. (2025). Review WHO (2016). Urban Green Spaces and Health (Version 1). figshare. https://doi.org/10.6084/m9.figshare.29526017.v1

8. Raina, A. S., Mone, V., Gorjian, M., Quek, F., Sueda, S., & Krishnamurthy, V. R. (2024). Blended physical-digital kinesthetic feedback for mixed reality-based conceptual design-in-context. In GI '24: Proceedings of the 50th Graphics Interface Conference (Article 6, pp. 1–16). ACM. https://doi.org/10.1145/3670947.3670967

9. Gorjian, M., Caffey, S. M., & Luhan, G. A. (2025). Exploring architectural design 3D reconstruction approaches through deep learning methods: A comprehensive survey. Athens Journal of Sciences, 12, 1–29. https://doi.org/10.30958/ajs.X-Y-Z

10. Li, J., V. Ossokina, I., A. Arentze, T., & Gorjian, M. (2025). Review of The Impact of Urban Green Space on Housing Value (Version 2). figshare. https://doi.org/10.6084/m9.figshare.29526167.v2